# Two-groove narrowband transmission filter integrated in a slab waveguide

Leonid L. Doskolovich,[1,2,*] Evgeni A. Bezus,[1,2] and Dmitry A. Bykov[1,2]

[1]Image Processing Systems Institute — Branch of the Federal Scientific Research Centre "Crystallography and Photonics"
of Russian Academy of Sciences, 151 Molodogvardeyskaya st., Samara 443001, Russia
[2]Samara National Research University, 34 Moskovskoe shosse, Samara 443086, Russia
*Corresponding author: leonid@smr.ru

We propose a simple integrated narrowband filter consisting of two grooves on the surface of a slab waveguide. Spectral filtering is performed in transmission at oblique incidence due to excitation of an eigenmode of the structure localized at a ridge cavity between the grooves. For the considered parameters, strictly zero reflectance and unity transmittance are achieved at resonant conditions. Width and location of the transmittance peak can be controlled by changing the widths of the grooves and of the ridge, respectively. The proposed filter may find application in waveguide-integrated spectrometers.

## 1. INTRODUCTION

Spectral and spatial filters that selectively transmit or reflect input light are indispensable in various optical devices including spectrum analyzers, image sensors, wavelength demultiplexers, liquid crystal displays and remote sensing systems. In a wide class of planar (integrated) optoelectronic systems, spectral or spatial filtering is performed in a slab waveguide [1–7]. Such geometry corresponds to the "insulator-on-insulator" platform and is suitable for the creation of fully integrated optical devices. In Refs. [1, 8–12], planar Bragg gratings (BGs) and phase-shifted Bragg gratings (PSBGs) for spectral filtering of optical radiation propagating along the waveguide were proposed. A planar BG corresponds to a single-mode slab waveguide with an appropriate periodically corrugated surface and acts as a reflection filter centered at the Bragg wavelength. A change in the waveguide thickness leads to a change in the effective refractive index of the guided mode and makes it possible to encode the required refractive index distribution analogous to a free-space BG. A planar PSBG consists of two or more symmetric BGs separated by defects (phase-shift regions) and enables to obtain a narrow transmission peak at the center of the stopband [1, 11, 12].

Recently, novel structures that can be applied in integrated spectrometers, the so-called digital planar holograms (DPHs), were proposed [3–6]. DPHs correspond to a sophisticated superposition of narrowband curvilinear BG-based reflectors fabricated on the surface of the waveguide layer. The curvilinear (e.g. elliptic) shape of the reflectors enables to perform spatially-resolved spectral filtering: radiation with different wavelengths is focused at different points. It is important to note that the methods for designing DPHs are a know-how of the authors and are not described in detail in the published works.

In this work, we propose a planar spectral filter consisting of two grooves on the surface of a slab waveguide and operating in transmission. In contrast to the PSBG-based filters, our filter has a much simpler geometry and is substantially more compact.

## 2. GEOMETRY OF THE STRUCTURE

The proposed filter is a planar version of dielectric three-layer resonant structure [Fig. 1(a)], which was studied in Ref. [13] for achieving large optical field enhancement over a large volume. A similar dielectric-metal-dielectric structure has been demonstrated to tunnel light through metal layers [14]. The structure in Fig. 1(a) consists of a central layer (refractive index $n_{wg}$, thickness $h_{wg}$) and two identical cladding layers (refractive index $n_{cl} < n_{wg}$, thickness $h_{cl}$).

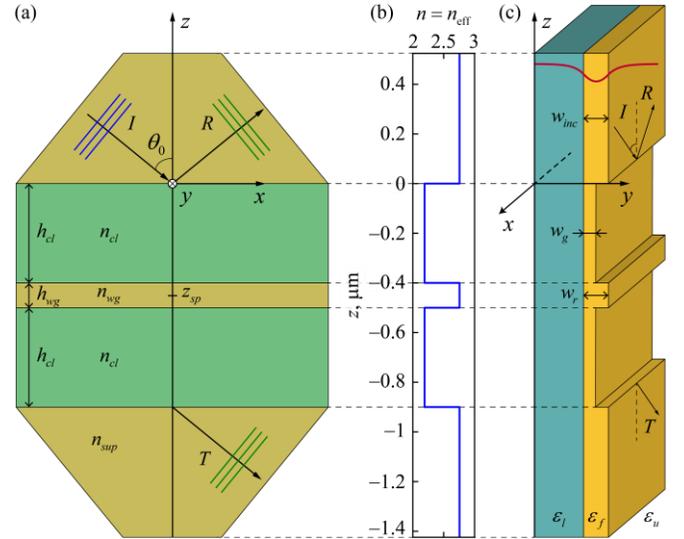

Fig. 1. Geometry of a conventional three-layer resonant structure (a) and of the proposed two-groove planar filter (c), and the refractive (effective refractive) index profile of the structure. The values in (b) and dimensions in (c) correspond to one of the examples considered below.

If the refractive index of the surrounding medium $n_{sup}$ satisfies the condition $n_{sup} > n_{cl}$, at a certain angle of incidence the structure may exhibit unity resonant transmission due to coupling of the incident light to a waveguide mode localized at the central layer, which in this case acts as a resonant cavity [13]. Let us note that the unity resonant transmission condition can be strictly satisfied only if the resonant structure has a horizontal symmetry plane $z = z_{sp}$ [Fig. 1(a)] [15, 16]. In this case, the structure corresponds to a so-called W-type waveguide [17]. At significantly large cladding thickness $h_{cl}$, the leaky mode of the W-type structure can be approximately described by the dispersion relation of the modes of a slab waveguide, which becomes exact at infinite $h_{cl}$. In this case, the excitation of the mode of the structure of Fig. 1(a) will occur in the vicinity of the frequency $\omega = \omega_0$ and the tangential component $k_x = k_{x,0}$ of the wave vector, which satisfy the dispersion relation of a slab waveguide [18]. Therefore, at resonant conditions the reflection coefficient will vanish at certain $\omega$ and $k_x$ values close to $\omega_0$ and $k_{x,0}$, respectively. It is worth noting that the quality factor of the resonance (the width of the resonant transmission peak) is determined by the thickness of the cladding layer $h_{cl}$ [13].

The presented analysis allows us to expect that the structure presented in Fig. 1(a) may be used as a spectral or spatial (angular) narrowband transmission filter. However, despite the simple geometry of the structure, it has several disadvantages concerning its practical use. In particular, due to the condition $n_{sup} > n_{cl}$, such a filter requires the use of two high-index prisms for incoupling and outcoupling of the incident optical beam, and thus cannot be highly integrated.

At the same time, this structure can serve as a prototype for designing a simple planar filter integrated in a slab waveguide. The considered waveguide has the following parameters: dielectric permittivity and the thickness of the core layer $\varepsilon_f$ and $w_{inc}$, respectively, and the dielectric permittivities of the claddings $\varepsilon_u$ and $\varepsilon_l$. As a planar analogue of the three-layer structure of Fig. 1(a), we propose to use the structure consisting of two grooves on the waveguide surface shown in Fig. 1(c). The lengths of the grooves and of the ridge between them are equal to the thicknesses of the layers of the conventional structure for free-space radiation.

Let us denote by $w_g$ and $w_r$ the thicknesses of the waveguide in the groove and ridge regions, respectively ($w_g < w_r$). At a fixed frequency $\omega = \omega_0$, effective refractive index of the guided mode $n_{eff}$ is a function of the core layer thickness $w$. Thus, the proposed planar structure can be considered as a direct analogue of a conventional structure with the refractive indices $n_{sup} = n_{eff}(w_{inc})$, $n_{cl} = n_{eff}(w_g)$, $n_{wg} = n_{eff}(w_r)$ [Fig. 1(b)]. In all considered examples, $w_r = w_{inc}$, so $n_{wg} = n_{sup}$. Since the function $n_{eff}(w)$ is monotonically increasing [18], $n_{wg} = n_{sup} > n_{cl}$.

## 3. RESULTS AND DISCUSSION

Let us numerically study the performance of the proposed two-groove planar filter (TGPF). In all the simulations, the following parameters were used: free-space wavelength $\lambda_0 = 630$ nm, $\varepsilon_f = 3.32^2$ (GaP), $\varepsilon_l = 1.45^2$ (fused silica), $\varepsilon_u = 1$, $w_{inc} = 100$ nm. At these parameters, the considered slab waveguide is single-mode. The TGPF consists of two grooves with depth $w_{inc}/2$ (so that $w_g = w_{inc}/2 = 50$ nm) and a ridge between them. In the case of TE-polarization, the effective refractive index of the incident guided mode and the mode in the ridge region is $n_{eff}(w_{inc}) = n_{eff}(w_r) = 2.7563$, while the effective refractive index in the groove region amounts to $n_{eff}(w_g) = 2.19$.

The resonances occurring in the proposed TGPF are associated with the excitation of a leaky eigenmode localized at the cavity corresponding to the central ridge and propagating along the $x$ axis. In this case, this mode is close to a mode of a photonic wire waveguide. The quality factor of the resonance (the width of the resonant peak) is determined by the width of the grooves $h_{cl}$. To illustrate this, we calculated the angular and wavelength transmission spectra of the investigated structure at two groove widths $h_{cl} = 400$ nm and $h_{cl} = 600$ nm (Fig. 2). As shown below, at these groove widths the angular and spectral widths of the resonant peaks differ by approximately 10 times. The width of the ridge is fixed and equals $h_{wg} = 100$ nm. The spectra were calculated using an in-house implementation of the aperiodic rigorous coupled-wave analysis (aRCWA) technique [19]. RCWA, also called the Fourier modal method, is an established numerical technique for solving Maxwell's equations. The angular spectra in Fig. 2(a) were calculated at fixed wavelength $\lambda_0 = 630$ nm. The transmittance reaches unity at the angles of incidence $\theta_1 = 55.201°$ (at $h_{cl} = 400$ nm) and $\theta_2 = 55.226°$ (at $h_{cl} = 600$ nm). The full widths of the resonant transmittance peaks at half maximum (FWHM) amount to $\Delta_{\theta,1} = 0.091°$ and $\Delta_{\theta,2} = 0.0092°$ at $h_{cl} = 400$ nm and $h_{cl} = 600$ nm, respectively. The dependence of the FWHM of the resonant peak on the width $h_{cl}$ is shown inset. Let us note that the angles of incidence $\theta_{1,2}$ depend on the thickness of the waveguide in the groove regions $w_g$. With a decrease in $w_g$ (decrease in $n_{eff}(w_g)$) the angles $\theta_{1,2}$ also decrease, and at $w_g = 0$ ($n_{eff}(0) = 1$ in the groove regions) they approach 33°.

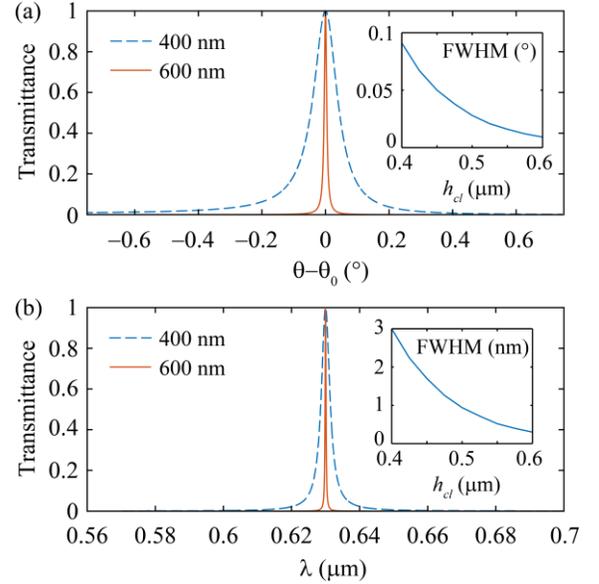

Fig. 2. Angular (a) and wavelength (b) TGPF transmission spectra at $h_{cl} = 400$ nm (dashed blue curves) and $h_{cl} = 600$ nm (solid red curves). The insets show the FWHM of the resonant peaks vs. $h_{cl}$.

It is worth noting that in the case of incidence of a TE-polarized mode at angles of incidence greater than $\theta > 46.34°$ there is no out-of-plane scattering and polarization conversion in the structure. Indeed, one can show that at $\theta > 46.34°$ the slab waveguide does not support any propagating TM-polarized modes with the propagation constants $k_{TM} > k_{x,TE}$, where $k_{x,TE} = k_0 n_{eff}(w_{inc}) \cdot \sin\theta$, $k_0 = 2\pi/\lambda_0$ is the tangential component of the wave vector of the incident TE-polarized mode. Moreover, the plane waves under and over the structure (in the regions with dielectric permittivities $\varepsilon_l$ and $\varepsilon_u$) with the $x$ wave vector component $k_{x,TE}$ are evanescent. Thus, the energy of the incident TE-polarized mode will be divided only between two modes (reflected and transmitted) having the same polarization. Due to this, strictly zero reflectance and unity transmittance are achieved in the structure at resonant conditions [15].

The wavelength spectra in Fig. 2(b) were calculated at the angles of incidence $\theta_1$ and $\theta_2$ taking into account the dielectric permittivity dispersion of the core layer. FWHM values of the resonant peaks are $\Delta_{\lambda,1} = 3.0$ nm (at $h_{cl} = 400$ nm) and $\Delta_{\lambda,2} = 0.3$ nm (at $h_{cl} = 600$ nm). Similarly to Fig. 2(a), the inset in Fig. 2(b) shows the FWHM of the peak vs. the width $h_{cl}$.

To further investigate the resonances associated with the transmittance peaks in Fig. 2, we calculated complex-wavelength eigenmodes of the TGPF corresponding to the poles of the transmission coefficient $T(\lambda)$ using the method of Ref. [20]. The obtained values $\lambda_{p,1} = 629.880 + 1.663i$ nm at $h_{cl} = 400$ nm and $\lambda_{p,2} = 629.932 + 0.162i$ nm at $h_{cl} = 600$ nm are in the vicinity of the wavelength $\lambda_0 = 630$ nm. Since the spectra in Fig. 2(b) have Lorentzian lineshape, the values $\Delta_1 = 2\operatorname{Im}(\lambda_{p,1}) = 3.326$ nm and $\Delta_2 = 2\operatorname{Im}(\lambda_{p,2}) = 0.324$ nm can be used to estimate the widths of the resonant peaks at half maximum. These values are in good agreement with the $\Delta_{\lambda,1}$ and $\Delta_{\lambda,2}$ values given above. Figure 3 shows the distributions of the $|H_y|^2$ component of the electromagnetic field at $h_{cl} = 400$ nm in the case of diffraction of the TE-polarized mode on the TGPF at three wavelengths: 600 nm (away from the resonance), 620 nm (closer to the resonance) and at 630 nm (at the resonance). The angle of incidence is fixed at $\theta_1 = 55.201°$. The distributions are normalized by the maximum intensity of the $H_y$ component of the incident guided mode. At the wavelengths 600 nm and 620 nm, the incident mode is almost completely reflected; the interference of the incident and reflected modes is clearly visible in Figs. 3(a) and 3(b). At 630 nm, a leaky eigenmode is excited at the ridge cavity between the grooves (Fig. 3(c)), and resonant unity transmittance is achieved. The incident and transmitted waves are not visible in Fig. 3(c) due to the field enhancement at the cavity. It is worth noting that the field distribution in Fig. 3(c) is visually identical to the field distribution of the leaky eigenmode of the structure.

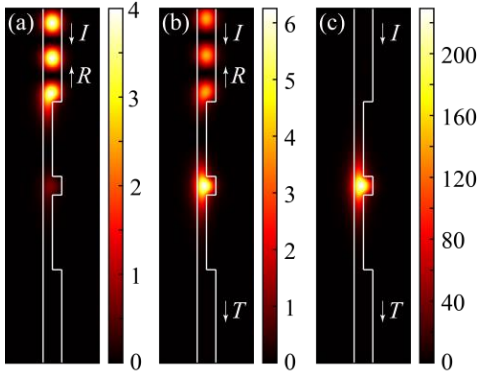

Fig. 3. Distributions of the $|H_y|^2$ component of the electromagnetic field in the TGPF normalized by the maximum $|H_y|^2$ value of the incident guided mode at $h_{cl} = 400$ nm and $\theta = \theta_1$ for the free-space wavelengths 600 nm (a), 620 nm (b) and 630 nm (c).

In the design of micro-spectrometers and hyperspectral sensors, so-called linear variable filters (LVFs) are widely used. LVF is a bandpass filter that is intentionally wedged in one direction, providing a position-dependent dispersive optical element. Typically, LVF corresponds to a Fabry–Pérot etalon with varying thickness between two Bragg reflectors [21–24]. On the basis of the proposed TGPF, a waveguide-integrated LVF can be created, in which the spectral shift of the transmittance peak can be achieved by changing one of the parameters of the structure, e.g. the ridge width $h_{wg}$. Figure 4 shows the possibility to control the location of the peak by changing the $h_{wg}$ value. In the considered example, the peak is shifted by approximately 7 nm if the $h_{wg}$ value is changed by 2.5 nm.

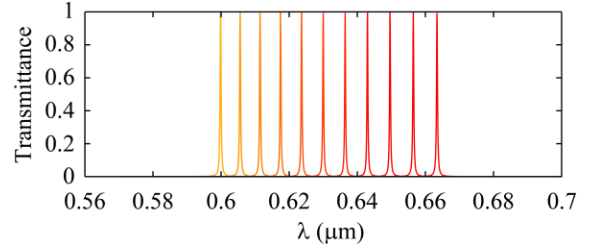

Fig. 4. Calculated transmittance spectra at varying ridge widths $h_{wg}$ changing from 87.5 nm (leftmost peak) to 112.5 nm (rightmost peak) in steps of 2.5 nm. Colors of the curves correspond to the resonant wavelengths at each ridge width value.

The given examples demonstrate that the proposed TGPF enables obtaining a narrow transmission peak with sub-nm spectral width and Lorentzian lineshape. In practical applications, achieving a nearly rectangular transmittance peak with a flat transmission band and steep slopes is highly desirable [11]. In PSBG-based filters, this problem is solved by introducing several properly chosen phase-shift regions. A similar approach can be applied for the considered TGPF.

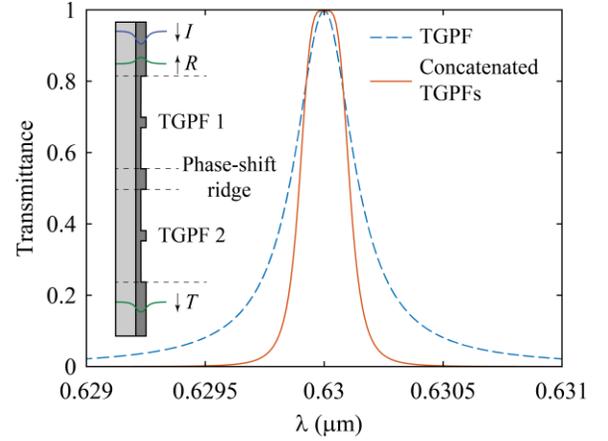

Fig. 5. Transmittance spectra of the TGPF with $h_{wg} = 100$ nm, $h_{cl} = 600$ nm (dashed blue curve) and of the planar structure consisting of two TGPFs separated by an additional groove with the width 198 nm (shown inset) (solid red curve).

As a simplest example, let us consider a filter composed of two concatenated TGPFs with $h_{wg} = 100$ nm and $h_{cl} = 600$ nm separated by an additional phase-shift ridge with the width 198 nm (inset in Fig. 5). Following the approach used in [25] for the analysis of reflectance spectra of stacked PSBGs, the width of the phase-shift ridge was chosen so that the reflectance spectrum of the structure possesses a so-called second-order zero. In terms of transmittance spectrum, this leads to a shape of the resonant peak that is significantly closer to a rectangle (Fig. 5).

The calculations [20] show that the concatenated structure supports two closely located modes with the complex wavelengths $\lambda_e = 629.896 + 0.082i$ nm (even mode) and $\lambda_o = 630.065 + 0.082i$ nm (odd mode). The even and odd modes of the concatenated structure correspond to the in-phase and out-of-phase excitation of the individual TGPFs' modes, respectively. Note that the complex wavelengths satisfy the condition $(\operatorname{Re}\lambda_o - \operatorname{Re}\lambda_e)^2 \approx 4\operatorname{Im}\lambda_o \cdot \operatorname{Im}\lambda_e$, which provides the so-called flat-top resonance line shape [26]. The quality factors of the modes of the concatenated structure are twice the quality factor of the mode of the initial structure ($\mathrm{Im}\,\lambda_o = \mathrm{Im}\,\lambda_e \approx \mathrm{Im}\,\lambda_{p,2}/2$). As the result, FWHM value for the concatenated structure decreases and in our example amounts to 0.22 nm instead of 0.3 nm for the initial structure. A further improvement in the peak shape can be achieved by increasing the number of concatenated TGPFs similarly to the approach used for PSBG-based filters [11].

## 4. CONCLUSION

In summary, in the present work we proposed a simple narrow-band transmission filter consisting of two grooves on the surface of a slab waveguide and intended for spectral filtering of optical radiation propagating in the waveguide. The spectral filtering is performed due to resonant excitation of an eigenmode of the structure localized at the cavity corresponding to the ridge between the grooves. On the basis of rigorous numerical simulations, the possibility to control the spectral width and location of the transmittance peak by changing the widths of the grooves and the ridge between them was demonstrated. Moreover, for the considered parameters, there is no unwanted parasitic scattering and polarization conversion in the filter. The proposed TGPF can be used for the creation of planar analogues of linear variable filters for waveguide-integrated spectrometers.

**Funding.** Russian Science Foundation (RSF) (14-19-00796) (investigation of the single TGPFs); Russian Foundation for Basic Research (RFBR) (16-29-11683, 15-07-00548) (investigation of the concatenated-TGPF structure and mode analysis).